\documentclass[longauth]{aa}  
\let\ACMmaketitle=\maketitle
\renewcommand{\maketitle}{\begingroup\let\footnote=\thanks \ACMmaketitle\endgroup}
\usepackage[english]{babel}
\usepackage[utf8]{inputenc}
\usepackage[T1]{fontenc}
\usepackage{graphicx}
\usepackage{amsmath,amssymb}
\usepackage{txfonts}
\usepackage[colorlinks=true,linkcolor=blue,citecolor=blue]{hyperref}
\usepackage{etoolbox}
\usepackage{wasysym}
\usepackage{threeparttable,adjustbox}
\usepackage{natbib}
\usepackage{xcolor}
\usepackage{placeins,afterpage}

\newcommand{\Msun}{$M_{\odot}$}
\newcommand{\Rsun}{$R_{\odot}$}
\newcommand{\Lsun}{$L_{\odot}$}

\newcommand{\kms}{km s$^{-1}$}

\newcommand{\Ha} {\mbox{H$\alpha$}~}
\newcommand{\Hb} {\mbox{H$\beta$}~}
\newcommand{\hei} {\ion{He}{i}~}

\newcommand{\Feii} {\ion{Fe}{ii}}
\newcommand{\Nai} {\ion{Na}{i} D }

\begin{document} 

\title{SN 2024abfo: a partially stripped SN~II from a yellow supergiant}

\titlerunning{SN 2024abfo}

\author{A. Reguitti\inst{1,2}\fnmsep\thanks{E-mail: andrea.reguitti@inaf.it},
A. Pastorello\inst{2},
S. J. Smartt\inst{3,4},
G. Valerin\inst{2},
G. Pignata\inst{5},
S. Campana\inst{1},
T.-W. Chen\inst{6},
A. Sankar. K.\inst{6}, \newline
S. Moran\inst{7}, 
P. A. Mazzali\inst{8},
J. Duarte\inst{9},
I. Salmaso\inst{10}, 
J. P. Anderson\inst{11,12},
C. Ashall\inst{13}, 
S. Benetti\inst{2},
M. Gromadzki\inst{14},\newline
C. P. Guti\'errez\inst{15,16},
C. Humina\inst{17}, 
C. Inserra\inst{18}, 
E. Kankare\inst{17}, 
T. Kravtsov\inst{19,17}, 
T. E. Muller-Bravo\inst{20,21},
P. J. Pessi\inst{22},\newline
J. Sollerman\inst{22}, 
D. R. Young\inst{4},
K. Chambers\inst{13}, 
T. de Boer\inst{13}, 
H. Gao\inst{13}, 
M. Huber\inst{13}, 
C.-C. Lin\inst{13}, 
T. Lowe\inst{13}, \newline
E. Magnier\inst{13}, 
P. Minguez\inst{13}, 
I. A. Smith\inst{13}, 
K. W. Smith\inst{3,4},
S. Srivastav\inst{3},
R. Wainscoat\inst{13},
M. Benedet\inst{23}\\
}

\authorrunning{Reguitti et al.} 

\institute{
INAF – Osservatorio Astronomico di Brera, Via E. Bianchi 46, I-23807 Merate (LC), Italy
\and
INAF – Osservatorio Astronomico di Padova, Vicolo dell'Osservatorio 5, I-35122 Padova, Italy
\and
Department of Physics, University of Oxford, Keble Road, Oxford, OX1 3RH, UK 
\and
Astrophysics Research Centre, School of Mathematics and Physics, Queen’s University Belfast, BT7 1NN, UK
\and
Instituto de Alta Investigación, Universidad de Tarapacá, Casilla 7D, Arica, Chile
\and
Graduate Institute of Astronomy, National Central University, 300 Jhongda Road, 32001 Jhongli, Taiwan
\and
School of Physics and Astronomy, University of Leicester, University Road, Leicester LE1 7RH, UK
\and
Astrophysics Research Institute, Liverpool John Moores University, ic2, 146 Brownlow Hill, Liverpool L3 5RF, UK
\and
CENTRA, Instituto Superior T\'ecnico, Universidade de Lisboa, Av. Rovisco Pais 1, 1049-001 Lisboa, Portugal
\and
INAF-Osservatorio Astronomico di Capodimonte, Salita Moiariello 16, I-80121, Naples, Italy
\and
European Southern Observatory, Alonso de C\'ordova 3107, Casilla 19, Santiago, Chile
\and
Millennium Institute of Astrophysics MAS, Nuncio Monse\~{n}or S\'{o}tero Sanz 100, Off. 104, Providencia, Santiago, Chile
\and
Institute for Astronomy, University of Hawai'i, 2680 Woodlawn Drive, Honolulu HI 96822, USA
\and
Astronomical Observatory, University of Warsaw, Al. Ujazdowskie 4, 00-478 Warszawa, Poland
\and
Institut d'Estudis Espacials de Catalunya (IEEC), Edifici RDIT, Campus UPC, 08860 Castelldefels (Barcelona), Spain
\and
Institute of Space Sciences (ICE, CSIC), Campus UAB, Carrer de Can Magrans, s/n, E-08193 Barcelona, Spain
\and
Department of Physics and Astronomy, University of Turku, Vesilinnantie 5, FI-20500, Finland
\and
Cardiff Hub for Astrophysics Research and Technology, School of Physics \& Astronomy, Cardiff University, Queens Buildings, The Parade, Cardiff, CF24 3AA, UK
\and
Finnish Centre for Astronomy with ESO, (FINCA), University of Turku, 20014 Turku, Finland
\and
School of Physics, Trinity College Dublin, The University of Dublin, Dublin 2, Ireland
\and
Instituto de Ciencias Exactas y Naturales (ICEN), Universidad Arturo Prat, Chile
\and
The Oskar Klein Centre, Department of Astronomy, Stockholm University, Albanova University Center, SE 106 91 Stockholm, Sweden
\and
Dipartimento di Fisica e Astronomia “G. Galilei”, Università di Padova, Vicolo dell’Osservatorio 3, I-35122 Padova, Italy
}

\date{Accepted 24 April 2025. Received March 2025; in original form November 2024}
 
\abstract{ 
We present photometric and spectroscopic data of the type IIb supernova (SN) 2024abfo in NGC~1493 (at 11 Mpc). 
The ATLAS survey discovered the object just a few hours after the explosion, and observed a fast rise on the first day.
Signs of the sharp shock break-out peak and the subsequent cooling phase are observed in the ultraviolet and the bluest optical bands in the first couple of days, while no peak is visible in the reddest filters. 
Subsequently, in analogy with normal SNe IIb, the light curve of SN~2024abfo rises again in all bands to the broad peak, with the maximum light reached around one month after the explosion. Its absolute magnitude at peak is $M_r=-16.5\pm0.1$ mag, making it a faint SN IIb.
The early spectra are dominated by Balmer lines with broad P-Cygni profiles indicating ejecta velocity of 22,500 \kms. One month after the explosion, the spectra display a transition towards being He-dominated, though the H lines do not completely disappear, supporting the classification of SN 2024abfo as a relatively H-rich SN IIb.
We identify the progenitor of SN 2024abfo in archival images of the Hubble Space Telescope, the Dark Energy Survey, and the XMM-Newton space telescope, in multiple optical filters. From its spectral energy distribution, the progenitor is consistent with being a yellow supergiant, having an initial mass of 15 \Msun.
This detection supports an emerging trend of SN IIb progenitors being more luminous and hotter than SN II ones, and being primaries of massive binaries. Within the SN IIb class, fainter events such as SN~2024abfo tend to have cooler and more expanded progenitors than luminous SNe IIb.
}


\maketitle

\section{Introduction}\label{introduction}

Type IIb Supernovae (SNe) are a type of core-collapse events from relatively massive and partially envelope-stripped stars \citep{Woosley1987, Filippenko1993}. During the first $\sim$30 days after the explosion, their spectra show strong and broad Balmer lines with P-Cygni profiles. Later, the H lines become weaker, while the \hei lines increase in intensity becoming the dominant spectral features. Hence, in a timescale of one month, SNe IIb evolve from H-dominated to He-dominated spectra
\citep[see][for a review]{Filippenko1997}.
The best-known SN~IIb is SN~1993J \citep{Wheeler1993, Richmond1994, Woosley1994, Barbon1995A&AS..110..513B}, although large samples of SNe IIb have also been studied \citep{Modjaz2014AJ....147...99M, Bianco2014ApJS..213...19B, Shivvers2019MNRAS.482.1545S, Pessi2019MNRAS.488.4239P, Rodriguez2023ApJ...955...71R}.

In a few notable cases, the progenitors of SNe IIb have been detected in archival images from the Hubble Space Telescope (HST), allowing the determination of their luminosities and colours, and therefore their masses and spectral types.
Their typical progenitors have been found to be relatively massive (10-18~\Msun) and extended ($R\simeq$200-600~\Rsun, see e.g. \citealt{Ouchi2017ApJ...840...90O, Yoon2017ApJ...840...10Y, Sravan2019ApJ...885..130S})  supergiant stars, with photospheric temperatures spanning a wide range of values (4200-11000)~K 
and luminosities in the range $\log L/{\rm L_{\odot}}\simeq4.7-5.1$
\citep{Crockett2008MNRAS.391L...5C, VanDyk2011ApJ...741L..28V, VanDyk2014AJ....147...37V, Kilpatrick2017MNRAS.465.4650K, Tartaglia2017ApJ...836L..12T, Niu2024ApJ...970L...9N}.

We present the study of the recently discovered SN~2024abfo, a type IIb SN in the nearby galaxy NGC~1493. We show that the progenitor is detected in archival images, including from HST, and it is compatible with the explosion of a yellow supergiant (YSG) star.

\section{Discovery and host galaxy}\label{discovery}

SN~2024abfo\footnote{also known as ATLAS24qew, PS24lge.} was discovered on 2024 November 15.9974 UT ($MJD = 60629.9974$) by the Asteroid Terrestrial-impact Last Alert System (ATLAS, \citealt{Tonry2018PASP..130f4505T}), through the Transient Science Server \citep{Smith2020PASP..132h5002S} at an ATLAS \textit{orange} ($o$) AB magnitude of $o = 18.55 \pm 0.18$. 
The SN exploded in the outskirts of the nearby galaxy NGC 1493 ($\sim$11 Mpc, \citealt{2024TNSAN.341....1S}), at celestial coordinates $\alpha =$ 03:57:25.61, $\delta = -$46:11:07.6. SN~2024abfo was classified on 2024 December 11 as a young type II SN by \cite{2024TNSAN.342....1W}.

We define as discovery epoch the MJD of the third image in a sequence of 30-second ATLAS exposures resulting in a 5$\sigma$ detection \citep[see][for a discussion on flagging discoveries in the ATLAS difference images, and the requirement of three 5$\sigma$ detections]{Smith2020PASP..132h5002S}. The source, detected with the ATLAS South African unit, was vetted by humans and registered on the Transient Name Server 12.7\,hours later at MJD=60630.5264 \citep{2024TNSTR4513....1T,2024TNSAN.341....1S}. 

The ATLAS Transient Server automatically runs forced photometry at the positions of candidate extragalactic transients, and this service is also publicly available \citep{2021TNSAN...7....1S}. The forced photometry for SN~2024abfo indicates further detections on the night before the formal discovery. Four exposures were obtained on 2024 November 14, taken with the ATLAS Chile unit. The first two images show no flux from the transient, neither through forced point-spread-function measurements nor in a visual inspection. The second two exposures indicate a rapidly rising supernova, suggesting we have caught the shock breakout (SBO) within the hour spanning the four ATLAS exposures. 
The forced flux measurements show that some supernova signal is detected between the two exposures taken on MJD 60628.28227 and 60628.29940 (hence, separated by 24.7 minutes). As observed in the early light curve of SN 2023ixf \citep{2024Natur.627..754L}, the flux during this very early phase rises faster than a simple fireball model ($f\propto(t-t_0)^2$). To estimate the time of first light, we adopt a broken power law fit \citep[Eqn. 1 of][]{2024Natur.627..754L} applied to the forced photometry on the eight individual 30 second exposures points on the nights of MJD 60628 and 60629, and find $t_0 = 60628.28 \pm 0.02$. We caution that between the two detections on MJD 60628 and  the next set of images 1.7 days later, there may well have been a strong shock breakout signature that we have not observed. 

The host galaxy NGC 1493 is a barred spiral with morphological classification as SB(r)cd \cite[according to][]{deVaucouleurs1991rc3..book.....D}. Its redshift is $z$ = 0.003512.
According to the NASA/IPAC Extragalactic Database (NED), the mean of the redshift-independent distance is $10.85 \pm 0.53$ Mpc ($\mu = 30.18 \pm 0.11$ mag), which we adopt in this work. These estimates come from Tully-Fisher method, such as from \cite{Bottinelli1984A&AS...56..381B, Bottinelli1985A&AS...59...43B, Bottinelli1986A&A...156..157B, Tully1988cng..book.....T}. The kinematic distances on NED are compatible, with values in the range $12\pm1$\,Mpc for $H_{0}=73$ \kms. 
The Galactic reddening towards SN 2024abfo is modest, at $A_V=0.03$ mag \citep{Schlafly2011ApJ...737..103S}, and the internal reddening is negligible (see Sect.~\ref{spectroscopy}), as we do not detect Na I D absorption lines in the SN spectra at the host galaxy redshift.

We estimated the metallicity at the location of SN 2024abfo by exploiting the statistical approach of \cite{Pilyugin2004A&A...425..849P} (their Equation 12). That relation between O/H and $M_B$ is valid at $0.4\times R_{25}$ ($83''$ for NGC 1493). SN 2024abfo is located $93''$\footnote{NED reports an axis ratio of 0.93, equivalent to an inclination of 21.5$^{\circ}$. If SN 2024abfo sits on the polar axis, the de-projected distance from the host centre turns to $100''$.} from the centre of the host galaxy, hence the approximation is fair. Using $B=11.88$ mag \citep{Lauberts1989spce.book.....L}, we obtain 12+log(O/H)=8.38$\pm$0.70, hence possibly a sub-Solar environment. Assuming 12+log(O/H)=8.69 dex for the Sun \citep{Asplund2021A&A...653A.141A}, we get $Z/Z_{\odot}$=0.5$^{+1.9}_{-0.4}$.

\section{Photometric evolution}\label{photometry}
\subsection{Ultraviolet and optical observations}
\begin{figure*}\centering
\includegraphics[width=1.36\columnwidth]{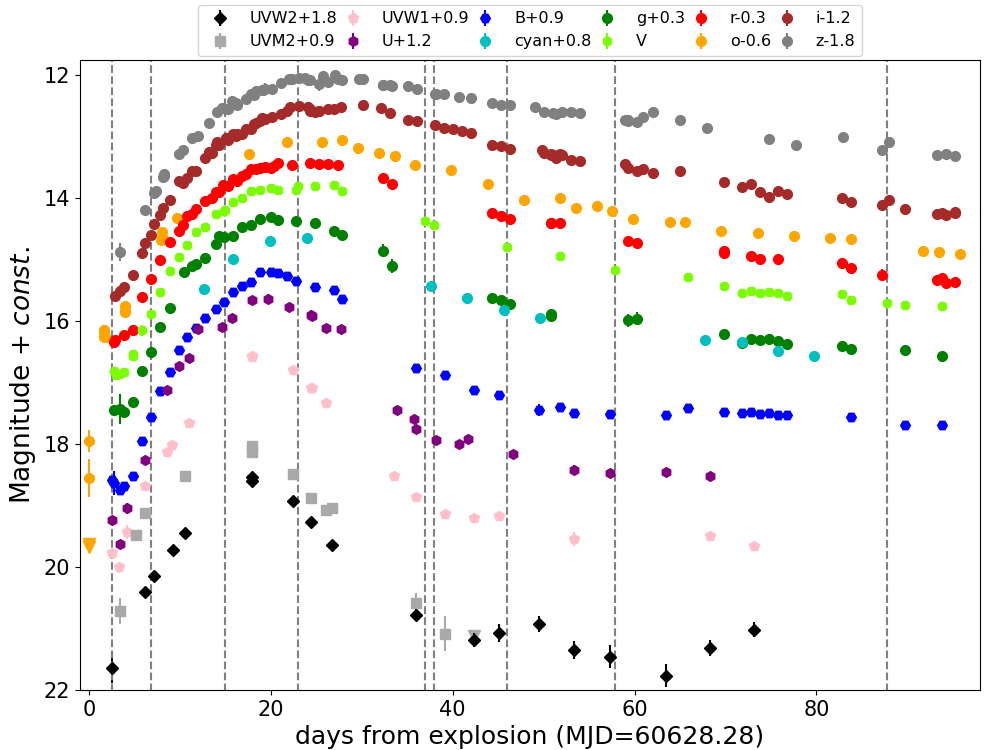}\includegraphics[width=0.68\columnwidth]{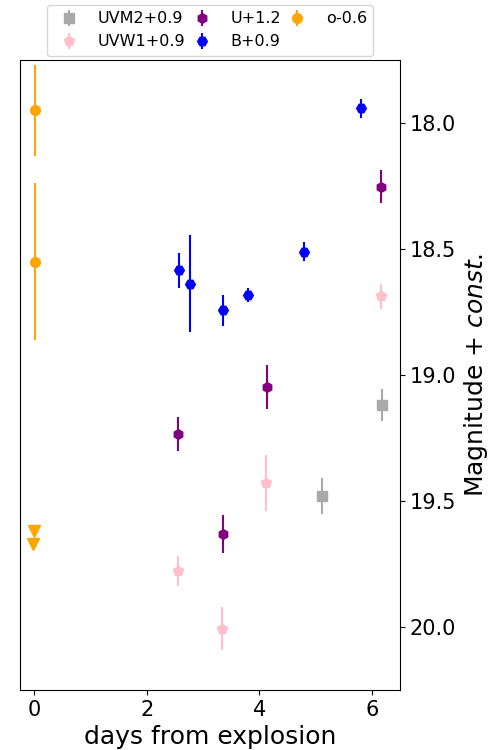}
\caption{Left: UV and optical light curves of SN 2024abfo in the first three months. Grey vertical dashed lines mark the epochs of the spectra.
Right: zoom on the early evolution in the ATLAS+UV/blue, highlighting the possible SBO and shock cooling.
}
\label{fig:light_curve}
\end{figure*}

SN 2024abfo was discovered very early, when it was still at an absolute magnitude of $M_{o} = -11.7 \pm 0.2$ mag, and was observed to rise rapidly soon after the discovery.
We started our multi-band follow-up a few days later, and the campaign lasted three months. We collected \textit{Swift} ultraviolet (UV) and optical photometric data with the facilities listed in Table \ref{tab1}.
The photometric data were reduced using standard procedures with the dedicated SNOoPY pipeline\footnote{{\sl ecsnoopy} is a package for SN photometry using PSF fitting and/or template subtraction developed by E. Cappellaro. A package description can be found at \url{http://sngroup.oapd.inaf.it/snoopy.html}.}. 
The \textit{Swift} UV data were reduced with the HEASOFT pipeline v.~6.32\footnote{NASA High Energy Astrophysics Science Archive Research Center – Heasarc 2014.}.
We also retrieve already calibrated measurements from the public surveys ATLAS and Pan-STARRS1 (PS1; \citealt{Chambers2016arXiv161205560C}).
The final UV and optical (Sloan, Johnson, and ATLAS) magnitudes are listed in the online supplementary materials, while the light curves are plotted in Figure \ref{fig:light_curve}, left panel.

After the fast rise observed by ATLAS, possibly linked to an SBO event \citep{Falk1977ApJS...33..515F, Nakar2010ApJ...725..904N, Waxman2017hsn..book..967W}, as described in Sect. \ref{discovery}, the bluer light curves slightly decline, while the redder bands magnitudes remain fairly constant (at $V\sim$16.8 mag). Instead, the UV data taken by \textit{Swift}/UVOT show a fast flux decline in the first two observations, which is clear evidence of SBO cooling (Fig. \ref{fig:light_curve}, right panel).
From +4 days, the light curves rise again in all filters (including those in the UV) towards the second, broader and brighter maximum, which is reached in the $r$-band $+24.1\pm0.1$ days from the explosion (obtained by fitting a 3rd-order polynomial on the data between $+10$ and $+30$ days).
In the UV filters, the maximum is reached between $+16$ ($UWM2$) and $+19$ days ($u$), while in the $i$ and $z$ bands it is reached 2 days after the $r$-band.

After maximum, the optical light curves decline linearly and slowly, whilst the UV light curves decline much faster.
In the first two months after the maximum, the $r$-band light curve declined by only $\sim$2 magnitudes, following a linear slope of $2.1\pm0.1$ mag (100 d)$^{-1}$.

\subsection{Bolometric light curve}
\begin{figure}
    \includegraphics[width=1\columnwidth]{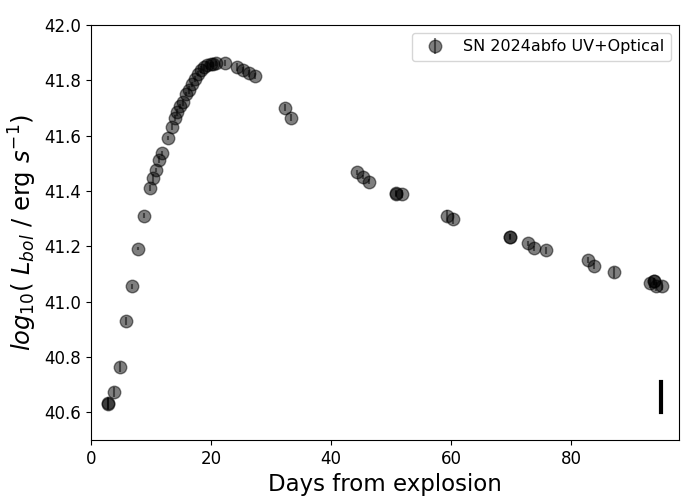}
    \caption{Pseudo-bolometric UV+Optical light curve of SN~2024abfo. The systematic error bar due to the uncertainty on the distance is reported in the bottom-right corner.}
    \label{fig:bolom}
    \end{figure}

We constructed the pseudo-bolometric light curve of SN~2024abfo from the contribution from $UVW2$ to $z$ bands. For epochs without observations in some bands, we interpolated the available data using the $r$-band light curve as a reference and assuming a constant colour index. The pseudo-bolometric light curve is shown in Fig. \ref{fig:bolom}.

We applied the simple “Arnett rule” \citep{Arnett1982ApJ...253..785A} to estimate key parameters of the SN explosion. For the ejecta mass $M_{\rm ej}$, we used Equation 1 of \cite{Prentice2019MNRAS.485.1559P}, adopting $\kappa=0.07$ cm$^{2}$~g$^{-1}$, $t_{\rm p}=21.6\pm0.4$ days (time between the explosion and the peak of the bolometric light curve), $v_{\rm ej}=8,000\pm500$ km s$^{-1}$ as measured from the position of the minimum of the P-Cygni profile of the \hei 5876 line in the +23 days spectrum (the closest to the peak) as a probe of the photospheric expansion velocity.
With these numbers, we estimate $M_{\rm ej}=4.1\pm0.7$~\Msun. The velocity of the \hei line is similar to that of the Fe lines, indicating that the line forms at the base of the He shell, as the results of non-thermal excitation \citep{Lucy1991ApJ...383..308L, Hachinger2012MNRAS.422...70H}. Therefore, the mass that is estimated with this method does not include the He shell and the H shell above it, both of which have low opacity, and should thus be increased by 1-2\,\Msun. 
Finally, the estimation of the kinetic energy using Equation 5 of \cite{Prentice2016MNRAS.458.2973P} is $E_k=(1.57\pm0.45)\times10^{51}$ erg.
The uncertainty inherent with the approach above is emphasised if the velocity of a H line is used. These probe the outermost region of the ejecta, which carry most of the $E_k$: if we use the velocity measured from the \Ha P-Cygni profile at +23 d as the proxy of the photospheric expansion velocity (13,100$\pm$400 \kms), we obtain: $M_{\rm ej}=6.7\pm0.7$~\Msun, $E_k=(6.7\pm1.1)\times10^{51}$ erg. This is also not a credible result, because the He and H shells have lower opacity than the metal-rich inner layers. The actual $E_k$ probably lies somewhere in between.
From the peak bolometric luminosity of $10^{41.87\pm0.12}$ erg s$^{-1}$ and $t_{\rm p}$, we can also provide an estimate of $M_{\rm 56Ni}$ using the same Arnett model and Equation 3 of \cite{Prentice2016MNRAS.458.2973P}. We obtain $M_{\rm 56Ni}=0.042\pm0.014$~\Msun, which is close to the value calculated for other SNe IIb (among those reported by \citealt{Prentice2016MNRAS.458.2973P}, only SN 2011hs is significantly lower).

The radioactive-powered part of the light curve starts to rise at +4 days, therefore, according to the \cite{Piro2021ApJ...909..209P} model\footnote{Given the small SBO, this model is not accurate at obtaining the progenitor radius (see Sect. \ref{progenitor}).}, $t_{ph}\approx4$ d. Using their same parameters and coefficients ($n=10$, $\delta=1.1$, $K=0.119$, $\kappa=0.2$ cm$^2$ g$^{-1}$ to account for the presence of He), and adopting $v_t=22,500$ \kms\,as measured from the H velocity in the first spectrum, we derive a mass of the H-rich external envelope roughly of the order of $M_e\approx0.08$ \Msun.
    
\subsection{Comparisons with similar objects}
\begin{figure}
\includegraphics[width=1\columnwidth]{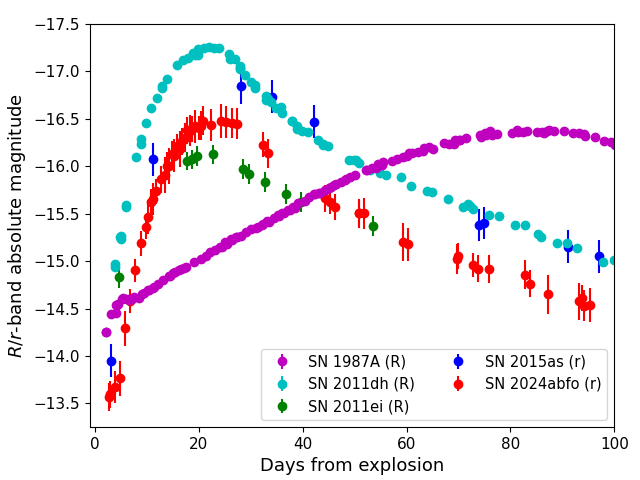}
\caption{$R/r$-band absolute light curve of SN 2024abfo compared with those of the faint type IIb SNe 2011dh \citep{Ergon2014A&A...562A..17E}, 2011ei, and 2015as, plus the peculiar type II SN 1987A \citep{1987MNRAS.227P..39M, 1987MNRAS.229P..15C}.
}
\label{fig:absolute}
\end{figure}

SN 2024abfo reaches an $r$-band peak absolute magnitude of $M_r=-16.5\pm0.1$ mag (Fig. \ref{fig:absolute}), making it a relatively underluminous SN IIb (as most SNe IIb peak in the range $-18.5 \leq M_R \leq -16.0$ mag, see \citealt{Morales-Garoffolo2014, Rodriguez2023ApJ...955...71R}).
Therefore, we select comparison objects among faint SNe IIb events, such as SNe 2011dh \citep{Marion2014ApJ...781...69M}, 2011ei \citep{Milisavljevic2013ApJ...767...71M}, and 2015as \citep{Gangopadhyay2018MNRAS.476.3611G}, plus SN 1987A - which is a canonical comparison object, though more H-rich. 

As can be seen in Fig. \ref{fig:absolute}, the peak absolute magnitude of SN~2024abfo in the $r$-band is $\sim$0.7 mag fainter than SN~2011dh, $\sim$0.4 mag dimmer than SN~2015as, but $\sim$0.3 mag brighter than SN~2011ei. Instead, its peak luminosity matches that of SN~1987A (in the $R$-band, \citealt{Hamuy1988AJ.....95...63H}). We note, however, that in SN 1987A the maximum is reached three months after the explosion, while in SN~2024abfo the peak is reached after less than one month.
The rise time of SN 2024abfo from the explosion to the maximum light in $r$-band is around 24~days, slightly longer than the average inferred from large samples of SNe IIb (21.3±2.1 days, \citealt{Pessi2019MNRAS.488.4239P}).
Finally, the post-peak light curve is parallel to that of SN 2011dh, i.e. they have the same linear decline slope.

\subsection{X-rays and radio observations}

\begin{figure}
\includegraphics[width=1\columnwidth]{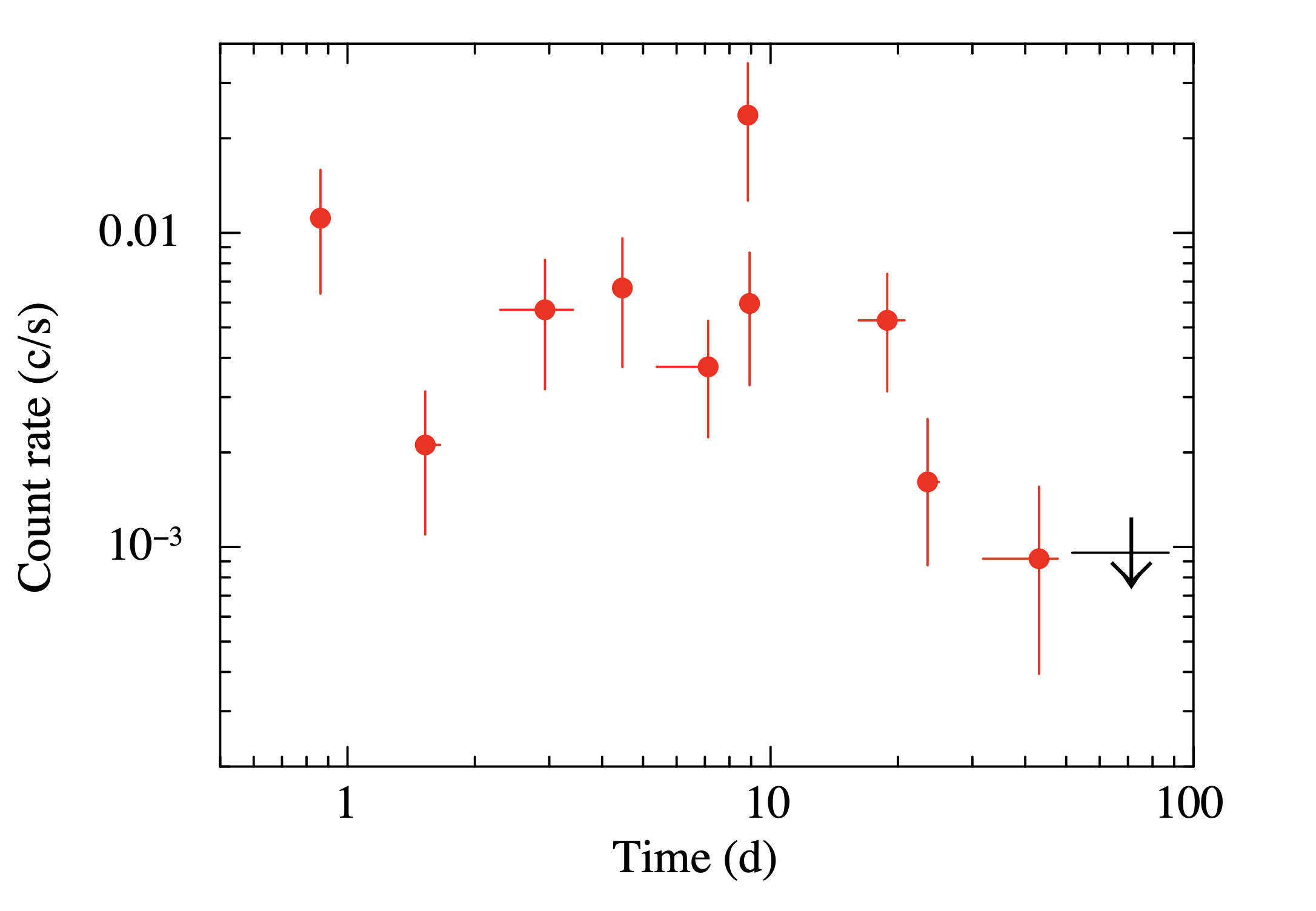}
\caption{X-ray light curve of SN 2024abfo, collected by {\it Swift}/XRT. The curve has been adaptively binned with a constant number of counts. The conversion factor for a power law model to a 0.3--10 keV unabsorbed flux is $6\times 10^{-11}$ erg cm$^{-2}$ counts$^{-1}$.}
\label{fig:Xcurve}
\end{figure}

SN 2024abfo was observed by \textit{Swift}/XRT on six occasions starting on 2024 November 16 at 20:26:10 UT. A weak source is detected in the first observations of 1.1 ks, with a count rate of $(8.0\pm3.7)\times 10^{-3}$ c~s$^{-1}$ (4 photons, signal-to-noise ratio 2.2). The source is then detected with an X-ray light curve tracking the optical one (see Figure \ref{fig:Xcurve}). It has to be noticed that in a previous {\it Chandra} image no sources appear in the nearby region surrounding SN 2024abfo. 
The X-ray observations point towards the detection of the SBO and cooling from SN 2024abfo also at high energies.

A reasonable spectrum can be obtained only for the entire data set, collecting $\sim 15$ source counts. Fixing the column density to the Galactic value of $\sim 10^{20}$ cm$^{-2}$ \citep{Willingale2013MNRAS.431..394W}, one derives a power law photon index of $1.9\pm0.8$ ($90\%$ confidence level) and an average 0.3--10 keV unabsorbed flux of $2\times 10^{-14}$ erg s$^{-1}$ cm$^{-2}$, or a bremsstrahlung temperature $k\,T=1.5^{+8.7}_{-0.8}$ keV and a factor $\sim 2$ lower flux.

We remark that SN 2024abfo was also detected at radio wavelengths by the Australia Telescope Compact Array at 9.0~GHz on 2024 November 21 \citep{ATel16920}. 
The flux density at 9.0~GHz is 0.36 ± 0.03 mJy, implying a luminosity of 5.2$\times10^{25}$ erg sec$^{-1}$ Hz$^{-1}$ at phase +7 d, while SN 2024abfo is not detected at 5.5~GHz to a 3-$\sigma$ upper limit of 0.06 mJy \citep{ATel16924}, suggesting the source was still in the early phase of evolution.

\section{Spectral evolution}\label{spectroscopy}

\begin{figure*}\centering
\sidecaption
\includegraphics[width=1.8\columnwidth]{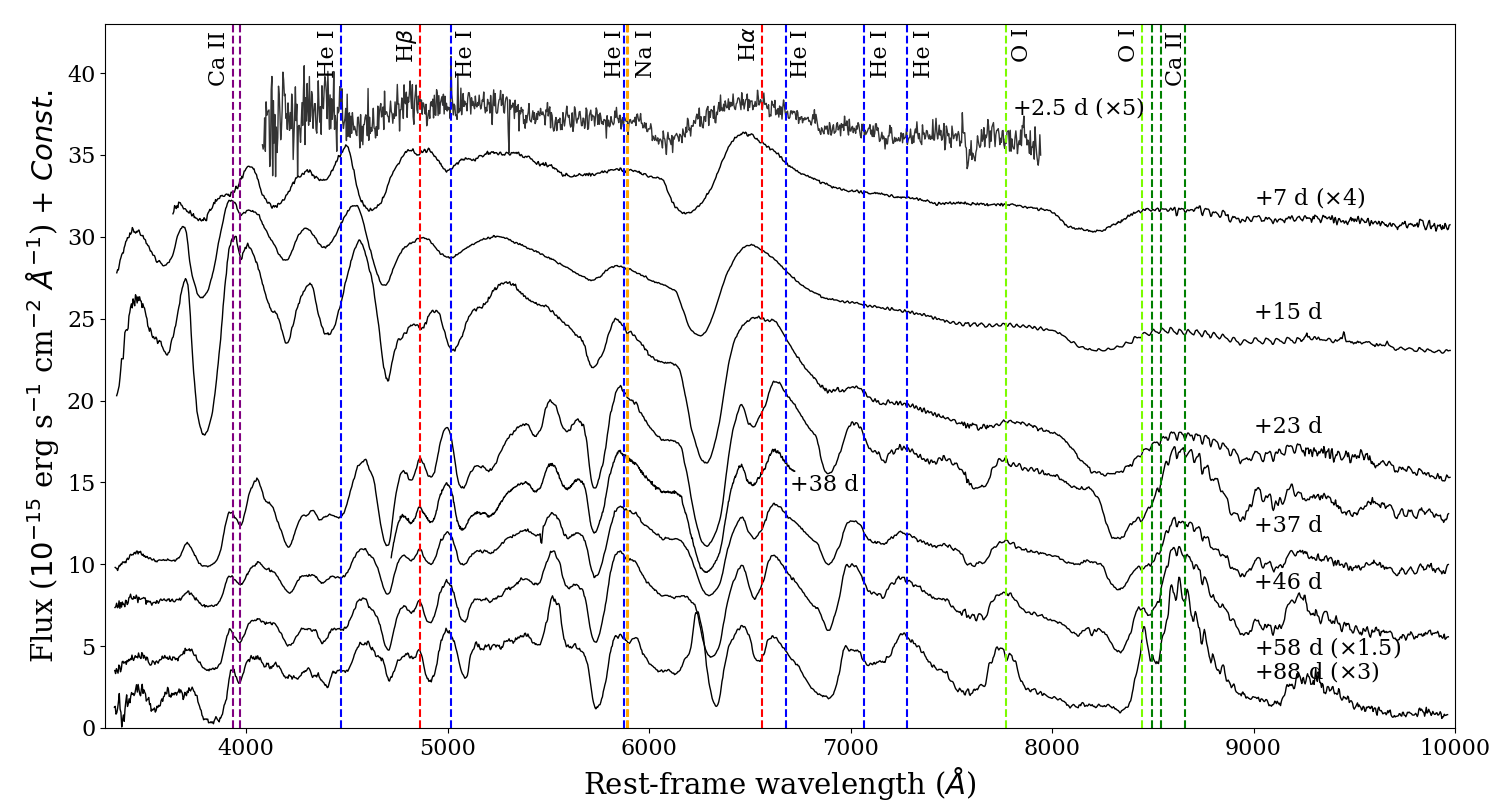}
\caption{Spectra of SN 2024abfo. The positions of the main lines are marked. The spectra are shifted by a constant for clarity.}
\label{fig:spectra}
\end{figure*}

SN 2024abfo was classified on 2024 November 16, 2.5 days after discovery, as a young type II SN \citep{Wet2024TNSAN.342....1W}.
In this first spectrum, shown in Figure \ref{fig:spectra}, the H$\alpha$ line has a broad P-Cygni profile. The expansion velocity measured from the position of the blue-shifted minimum of the broad H$\alpha$ P-Cygni is $v_{\rm H\alpha} = -22,500\pm1,000$ \kms; extremely fast even accounting for the very early phase at which the spectrum was taken.

We conducted a spectroscopic follow-up of SN~2024abfo, during which we collected nine optical spectra spanning the first three months of evolution. The log of spectroscopic observations is provided in Table \ref{tab:spectra}, and the spectral time series is presented in Figure \ref{fig:spectra}.
The spectra from the 3.58~m New Technology Telescope (NTT) equipped with \textsc{EFOSC2} were collected in the framework of the extended European Southern Observatory Spectroscopic Survey of Transient Objects (ePESSTO+) collaboration, and reduced with a \texttt{pyraf}-based pipeline (\texttt{PESSTO}), optimised for the \textsc{EFOSC2} instrument
\citep{2015A&A...579A..40S}. 

The spectrum at +7 days, shown in Figure \ref{fig:spectra}, still presents H$\alpha$ with a broad P-Cygni profile, from which we infer an ejecta velocity of $v_{\rm H\alpha} = -17,500\pm500$ \kms. The P-Cygni profile is now visible also in the \Hb and Ca II NIR lines, while a trapezoidal emission from \hei 5876 or Na I D also appears. 
In the +23 d spectrum, the main \hei emission lines start to appear. In particular, the blend of \hei $\lambda$6678 with the \Ha line produces a broad flat-topped feature. 
In the +37 d spectrum, the continuum is redder due to the blanketing from metal lines, while a very broad P-Cygni profile due to the Ca II NIR triplet also emerges in the reddest region of the spectrum. While \hei $\lambda$7065 is now clearly detected, both \hei $\lambda$5876 and $\lambda$6678 become stronger than H$\alpha$, while \hei $\lambda$5015 is more prominent than H$\beta$, though the H lines still remain visible. 
Through the measure of the minimum of the P-Cygni profile of H$\alpha$, we note a clear declining trend of the ejecta velocity as time progresses, with $v_{\rm H\alpha}$ decreased to 10,000~\kms\,in the +37 d spectrum.
Although the +38 d spectrum has a higher resolution, the narrow \Nai absorption lines at the rest wavelength and at the redshift of NGC~1493 are not detected, hence, according to the \cite{Poznanski2012MNRAS.426.1465P} relation, confirming the minimal Galactic and host galaxy extinction contributions.
We measured the evolution of the velocities of the H$\alpha$, \hei $\lambda$5876 and \Feii~$\lambda$5169 lines in all the spectra, from the position of the minimum of the respective P-Cygni profiles. The values are reported in Table \ref{tab:vel_spectra}.

In the +88 d spectrum, some nebular transitions also start to appear, like the O I $\lambda$7774 and $\lambda$8446. The [O I] $\lambda$6300 is also present in emission. The \hei $\lambda$7281 line is now stronger than the $\lambda$7065 one, which suggests a contamination from the [Ca~II] $\lambda\lambda$ 7291,7324 doublet. The \Ha P-Cygni is narrower than in the +58~d spectrum. From its minimum, we derive a velocity of $v_{\rm H\alpha} = -10,500\pm500$ \kms.

\section{Progenitor detection}\label{progenitor}

\begin{figure*}\centering
\includegraphics[width=2.25\columnwidth]{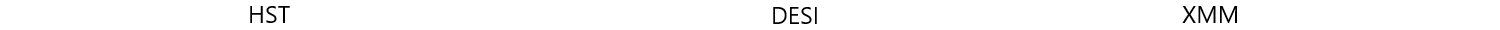}\\
\includegraphics[width=.75\columnwidth]{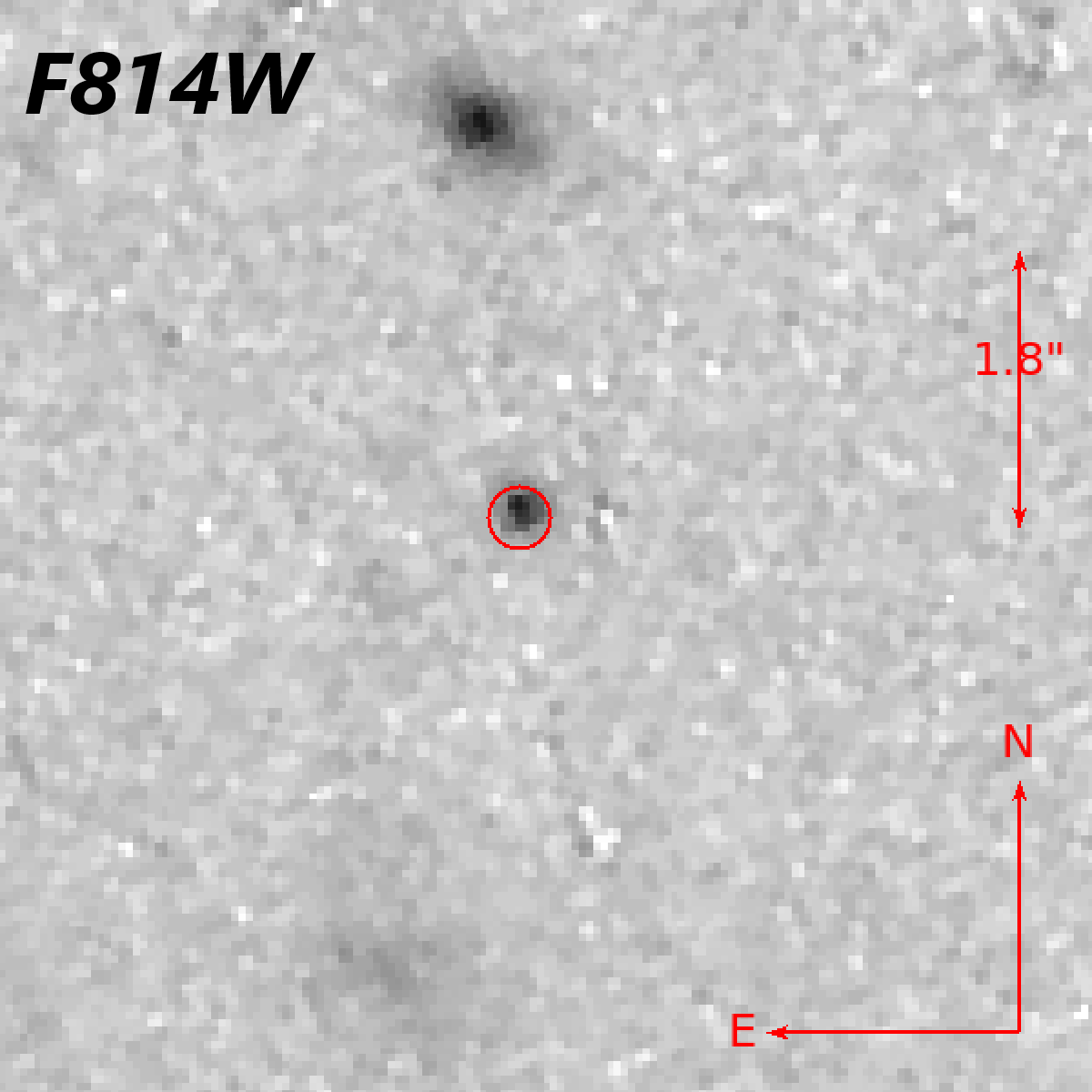}\hspace{0.5em}
\includegraphics[height=.75\columnwidth]{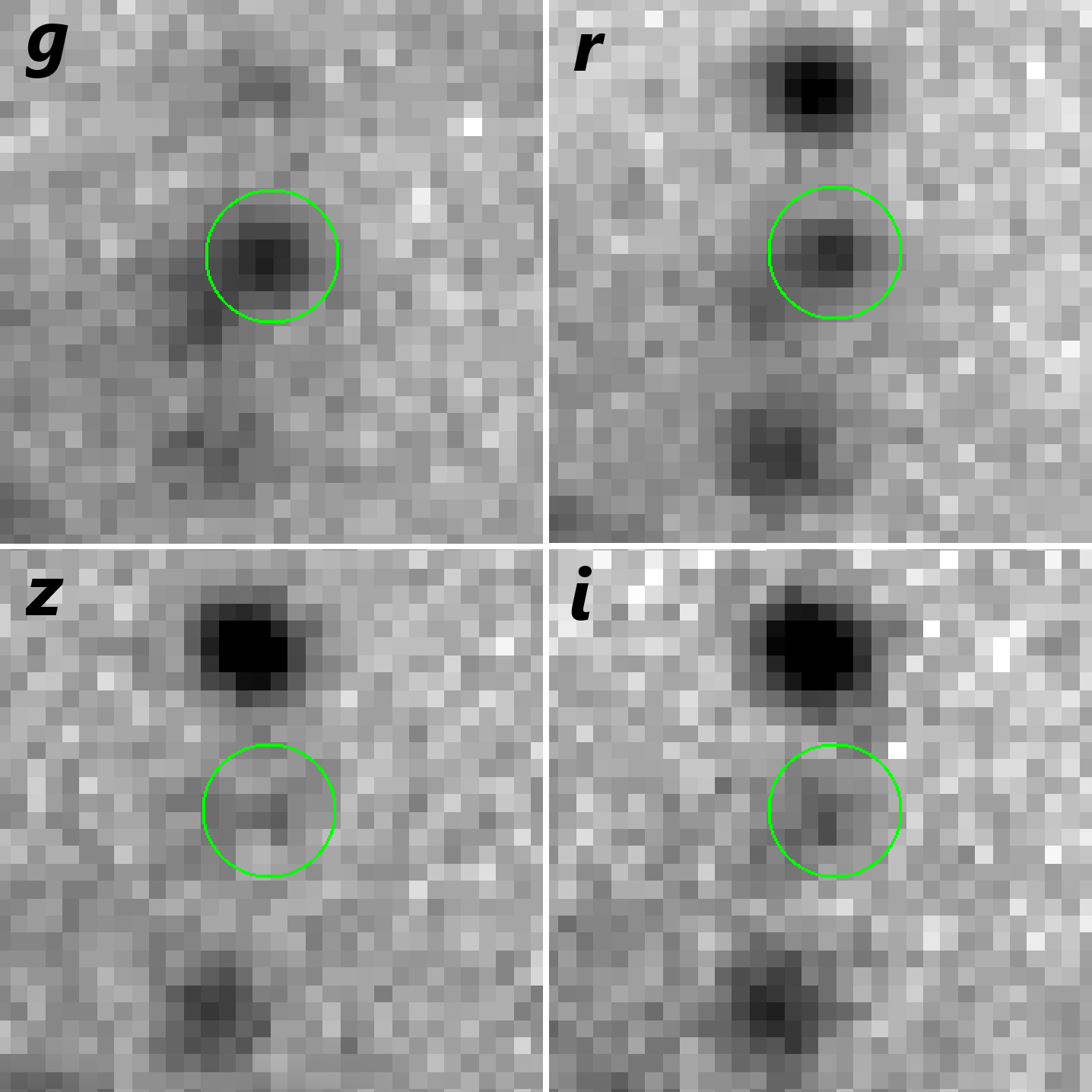}\hspace{0.5em}
\includegraphics[height=.75\columnwidth]{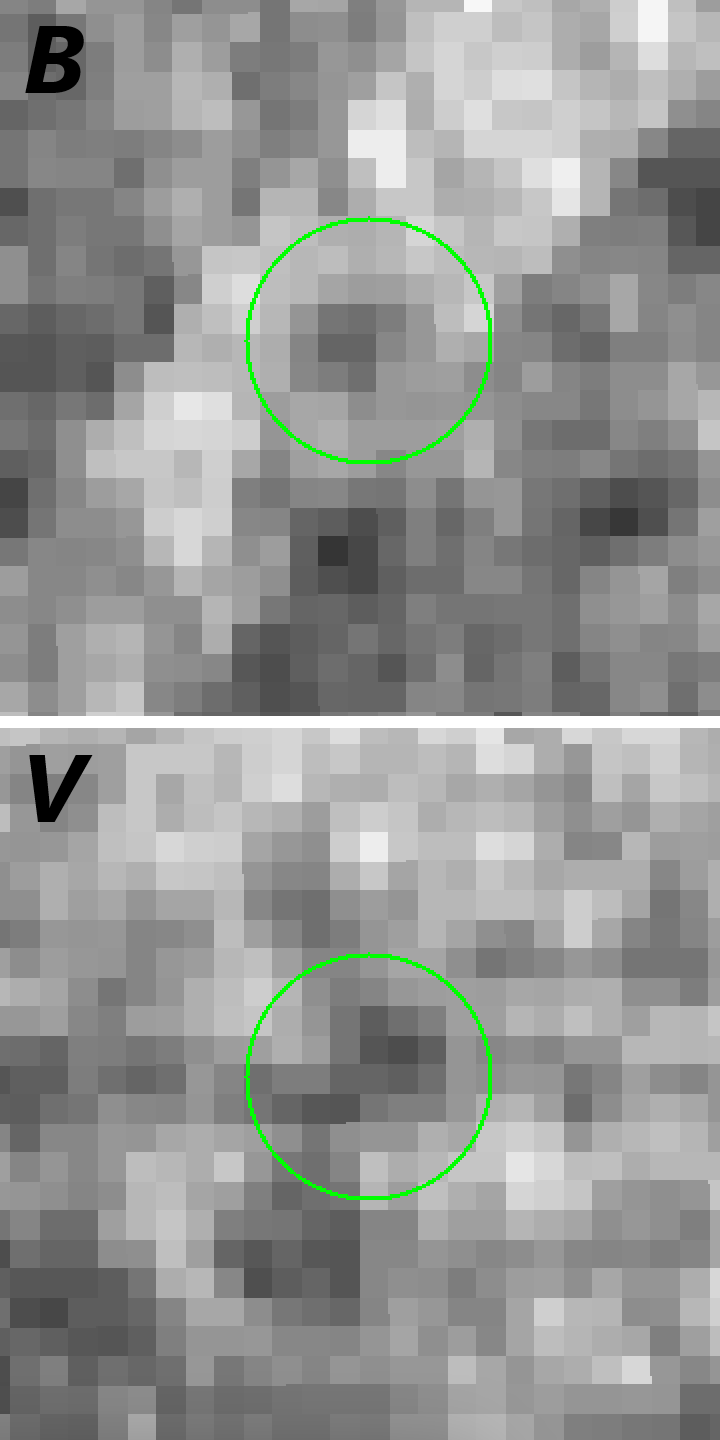}
\caption{
Left: the progenitor of SN 2024abfo detected in an HST/WFPC2 \textit{F814W} image obtained in 2001. The circle is centred at the transformed position of SN 2024abfo, and has a radius equal to the uncertainty SN position of $0.2''$.
Centre: the progenitor of SN 2024abfo is visible in the stacked images from DESI Legacy Surveys DR10 taken between 2013 and 2017. In a clockwise sense from the upper left corner: Sloan $g$, $r$, $i$, $z$ images. 
Right: the progenitor detected also by \textit{XMM-Newton/OM} in 2010 in $B$ (top) and $V$ (bottom) filters. 
The circles are centred at the position of SN 2024abfo; they have a radius of $1''$ in the DESI frames and $2''$ in the XMM frames. In all stamps, North is up and East to the left.}
\label{fig:progenitor}
\end{figure*}

We analysed an \textit{HST/WFPC2} image taken on 2001 May 02 with the \textit{F814W} filter (Proposal ID 8599, PI Boeker). 
To identify the SN position in the \textit{HST} image, we used a $z$-band image ($0.4''$ pix$^{-1}$, $26'\times26'$ field of view; exposure 200 s) obtained on 2024 November 22.55 UT at Las Cumbres Observatory (LCO), under seeing $1.7''$.
We achieved good-precision relative astrometry between the \textit{HST} pre-explosion images and the LCO post-explosion image, by geometrically transforming the former to the latter. We registered five point-like sources in common between the two datasets. Then, using \texttt{geomap} and \texttt{geoxytran} (two \texttt{IRAF} tasks), we carried out a geometrical transformation between the sets of coordinates. The uncertainty in the SN position is calculated as a quadrature sum of the $rms$ of the geometric transformation ($0.16''$), and the error on the SN centroid position in the LCO image ($0.12''$).
The progenitor star is clearly visible at the same position of SN 2024abfo, at pixel coordinates [2489.2, 3643.3], with \textit{F814W} = 22.32 ± 0.05 AB mag (Fig. \ref{fig:progenitor}, left). Using the adopted distance to NGC 1493, it corresponds to an absolute $M_{\rm F814W}$ = $-$7.9 ± 0.1 mag. 
The same source is also detected in the stacked frames between 2013 and 2017 from the DESI Legacy Imaging Surveys \citep{DESI2016arXiv161100036D, Dey2019AJ....157..168D} DR10, with catalogue\footnote{https://www.legacysurvey.org/viewer/ls-dr10/cat?ralo=59.3559\&rahi=59.3579\&declo=-46.1865\&dechi=-46.1845\&objid=9754} Sloan filters magnitudes $g$ = 22.32, $r$ = 22.21, $i$~=~22.12, $z$ = 22.64 AB mag (Fig.~\ref{fig:progenitor}, centre). We assume an error of 0.1 mag on these values.
The progenitor is also detected in images taken by \textit{XMM-Newton/OM} on 2010 July 15 (PI Mathur), at $B$ = 22.66 ± 0.23, $V$ = 22.44 ± 0.06 Vega mag (Fig. \ref{fig:progenitor}, right).

\begin{figure}\centering
\includegraphics[width=\columnwidth]{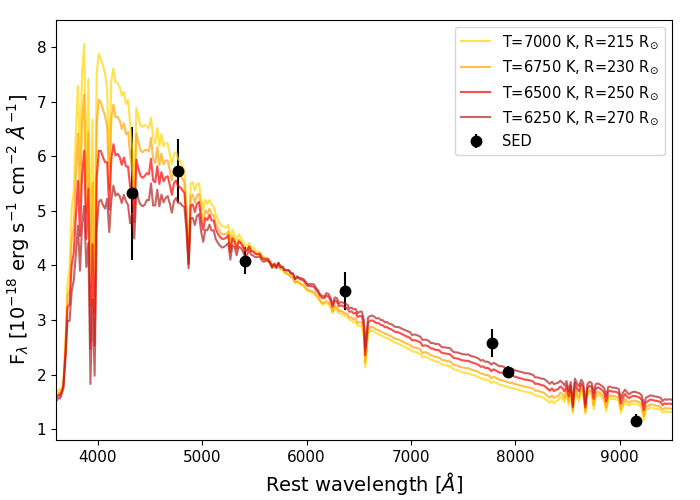}
\caption{SED of the progenitor of SN 2024abfo compared to ATLAS9 stellar atmospheric models from \cite{Kurucz1993}. Good matches are found with models with temperatures between 6250 and 7000 K, and radii between 215 and 270 \Rsun.}
\label{fig:SED_progenitor_Kurucz}
\end{figure}

\begin{figure}\centering
\includegraphics[width=1.02\columnwidth]{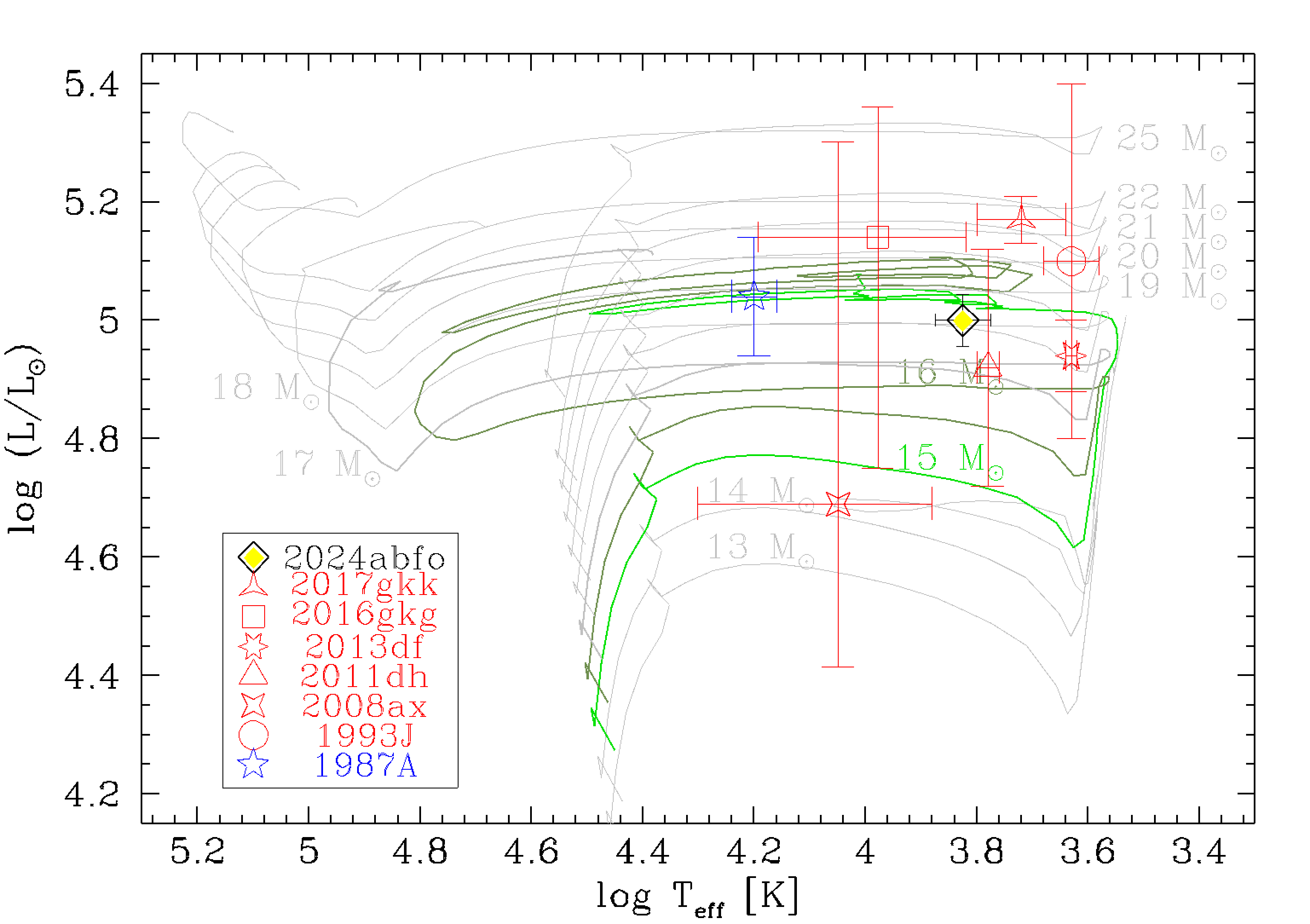}
\caption{Location of the progenitor of SN 2024abfo within the HRD from the match of ATLAS9 models to its SED. BPASS evolutionary tracks for binaries between 13 and 25~\Msun\,are also plotted for comparison. The progenitor is consistent with a star with an initial mass of 15 \Msun. The location of the other identified progenitors of SNe IIb and of SN 1987A are also shown.}
\label{fig:HRD}
\end{figure}

These colours and the inferred absolute magnitudes suggest a relatively blue and luminous progenitor. 
We constructed the spectral energy distribution (SED) by converting the observed magnitudes into flux densities at the effective wavelengths of each filter, after correcting for the small interstellar extinction.
We compared the SED of the progenitor of SN 2024abfo with the ATLAS9 stellar atmospheric models by \cite{Kurucz1979ApJS...40....1K, Kurucz1993}\footnote{retrieved here: \url{https://archive.stsci.edu/hlsps/reference-atlases/cdbs/grid/k93models/}}. We adopted the models with $Z/Z_{\odot}=0.5$ (hence log$(Z) =-0.3$) and log$(g) =1.0$. 
We found good matches for models with photospheric temperatures between 6250 and 7000 K, and stellar radii between 215 and 270 \Rsun\,(for the hottest and coldest model, respectively). Their corresponding bolometric luminosities are $(3.83\pm0.04)\times10^{38}$ erg~s$^{-1}$ (hence, log($L/L_{\odot})=5.00\pm0.05$ (accounting for the error on the distance modulus).
These parameters correspond to a YSG of late F spectral type. The matches of the ATLAS9 models to the progenitor's SED are shown in Fig. \ref{fig:SED_progenitor_Kurucz}. Its location in the Hertzsprung–Russell diagram (HRD) is shown in Fig. \ref{fig:HRD}.
We also over-plotted BPASS (v. 2.3) evolutionary tracks \citep{Byrne2022MNRAS.512.5329B, Stanway2018MNRAS.479...75S, Eldridge2017PASA...34...58E} for binary systems with a metallicity of $Z=0.01$, a secondary vs. primary mass ratio $q=0.1$, and a binary orbital period of $P=1000$~days. We chose these parameters as illustrative of a detached system.
We favour a binary progenitor based on the observed SED (rather blue, indicative of envelope stripping), the fact that a majority of massive stars are found in binary systems \citep{Sana2012Sci...337..444S}, and the detection of the surviving companion in stripped-envelope SNe (for instance, \citealt{Maund2004Natur.427..129M, Fox2022ApJ...929L..15F}). The tracks shown in the figure are for primaries with initial masses between 13 and 25~\Msun. The comparison allows us to estimate for the progenitor of SN~2024abfo an initial mass of 15 \Msun.
The progenitor of SN 2024abfo is among the hottest progenitors identified or derived for SNe IIb, with a photospheric temperature of 6250-7000~K (or log$(T_{\rm eff})=3.80-3.85$), close to that found for the progenitors of SN~2016gkg (9550~K, \citealt{Kilpatrick2017MNRAS.465.4650K}) and SN~2008ax (7900-20000~K, \citealt{Folatelli2015ApJ...811..147F}).

The source at the position of SN~2024abfo in the HST/WFPC2 images appears to be very isolated. Although the DESI images have low spatial resolution ($0.26''$), the HST/WFPC2 images indicate no strong contamination to the single PSF fit that describes the source. This is only true for the $F814W$ and DESI $i$-band images. The bluer filters of DESI show nearby extended flux to the south-east. While the evidence of a point-like source suggests we are measuring the progenitor system rather than a cluster or stellar blends, we caution that higher resolution imaging of SNe IIb progenitors in the years after explosion shows that the luminosity can be overestimated \citep[e.g.][]{Folatelli2015ApJ...811..147F, 2022ApJ...936..111K}, although in some cases it is unaffected, as in the case of SN 2011dh \citep{2015MNRAS.454.2580M}. 

Additionally, the progenitor is likely a binary system, which could be composed of a slightly cooler supergiant and a fainter, but hotter and more compact companion (like the system of SN~1993J, \citealt{Maund2004Natur.427..129M}), and we are measuring only the integrated flux. If the mass ratio $q$ is indeed 0.1, it means the secondary star would have a mass of only $\sim$1.5 \Msun, which does not contribute significantly to the luminosity of the system.

\section{Discussion and conclusion}\label{discussion}

In the spectra of normal SNe IIb, the \hei lines become the most prominent features about one month past explosion, while the H lines almost disappear. In contrast, while in the +29 d spectrum of SN~2024abfo the He lines are dominant, \Ha and \Hb are still present, suggesting more H-rich ejecta. Therefore, while the progenitor star had likely lost a large fraction of its outer envelope before the core collapse, a significant amount of H was still present. 
SN 2024abfo is a rare case in which the shock breakout may have been caught during the discovery, although it is very faint, only noticeable in the early ATLAS data.

According to \cite{Chevalier2010ApJ...711L..40C}, those SNe IIb showing an early, sharp light-curve peak due to SBO cooling before rising to the main $^{56}$Ni-decay-powered maximum (e.g., SN~1993J, \citealt{Wheeler1993}), arise from the explosion of extended stars, whereas those lacking such a feature originate from more compact objects.
Compact SNe IIb progenitors could have been evolved, post-Wolf-Rayet stars \citep[][and references therein]{Soderberg2006ApJ...651.1005S}; after the detection the progenitor of SN~2008ax, they were suggested to be quite massive ($\approx$10–28~\Msun; \citealt{Crockett2008MNRAS.391L...5C}). 
\cite{Folatelli2015ApJ...811..147F} showed that the SN~2008ax pre-explosion progenitor source is not single and that a massive interacting binary system was more consistent with the data (an 18 \Msun+12 \Msun\,on a $P^{i}_{\rm orb}=5$ days orbit). Radially extended progenitor configurations may suggest lower-mass single stars or binary progenitor systems, with their outer H envelopes being partially stripped by strong stellar winds or through interaction with the companion, respectively \citep{Podsiadlowski1992ApJ...391..246P, Folatelli2014ApJ...792....7F, Fox2014ApJ...790...17F, Ryder2018ApJ...856...83R}.

\begin{figure}\includegraphics[width=1\columnwidth]{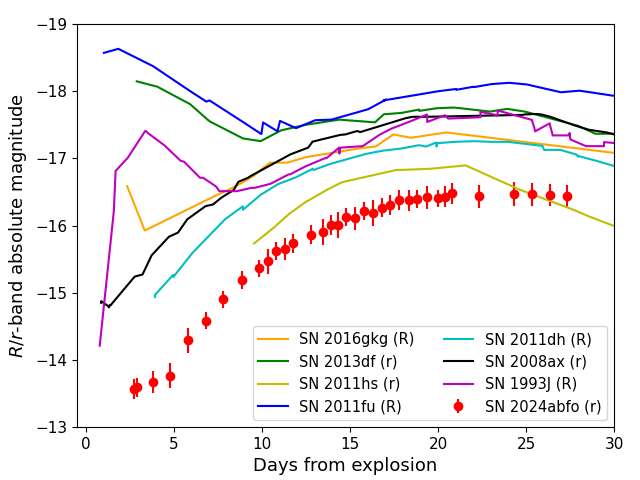}
\caption{$R/r$-band early absolute light curves of a small SNe~IIb sample with well-constrained progenitor parameters (see the text). 
}
\label{fig:absolute_all}
\end{figure}

We now consider the small sample of SNe IIb with detected progenitors (SNe 1993J, 2008ax, 2011dh, 2013df, 2016gkg, 2017gkk, and 2024abfo; \citealt{1994AJ....107..662A, Crockett2008MNRAS.391L...5C, Maund2011ApJ...739L..37M, VanDyk2014AJ....147...37V, Folatelli2015ApJ...811..147F, Tartaglia2017ApJ...836L..12T, Niu2024ApJ...970L...9N}, this work), or whose properties were determined via hydrodynamical modelling of the bolometric light curve (for instance SNe 2011fu and 2011hs, \citealt{Morales-Garoffolo2015, Bufano2014MNRAS.439.1807B}). We note a general trend (Fig. \ref{fig:absolute_all}): SNe IIb that are fainter at peak (such as SNe 2011dh\footnote{but see the $g$-band light curve from \cite{Arcavi2011ApJ...742L..18A}.}, 2011ei, 2015as) usually do not show (or is faint, like in SN 2024abfo) an early, short shock-cooling peak in their light curves. Instead, this feature is clearly present in the most luminous events, like in SNe 2011fu and 2013df, \citealt{Morales-Garoffolo2014, Morales-Garoffolo2015}).
At the same time, as highlighted in Tab. \ref{tab:progenitors}, more luminous SNe IIb (often showing the first peak) have cooler and more extended progenitors, while lower luminosity objects (often without the first peak) have hotter and more compact progenitors. 
If we extend the sample to other faint SNe IIb, such as SN 2011ei and 2015as (already shown in Fig. \ref{fig:absolute}), this trend becomes even more pronounced.
SN 2016gkg is somewhat an exception, given it has a hot (9550~K) and less extended (250~\Rsun, \citealt{Kilpatrick2017MNRAS.465.4650K}) progenitor while presenting a first, early light-curve peak \citep{Bersten2018Natur.554..497B}.

Interestingly, most SN IIb progenitors have a luminosity in the log($L/L_{\odot}$)$=4.9-5.1$ range, higher than normal SNe II \citep[4.2-4.7;][]{Smartt2009ARA&A..47...63S}. SN 2024abfo is another object that supports this emerging trend. This is unlikely to be an observational bias, as the smaller bolometric correction associated to a white/yellow supergiant star should allow us to detect them even at lower bolometric luminosities compared to the red supergiant progenitors of typical SNe IIP. In addition, there is growing evidence that the progenitors of SNe IIb are preferentially found in high-mass binaries, while normal SNe II are the outcome of single-star evolution.
A continuum in the progenitor properties seems to exist from the most H-rich red supergiant stars (producing normal type II/IIP SNe, e.g. \citealt{Smartt2009MNRAS.395.1409S}), the moderately H-rich yellow hypergiant precursors of SNe IIL \citep{Elias-Rosa2010ApJ...714L.254E, Elias-Rosa2011ApJ...742....6E, Fraser2010ApJ...714L.280F}, SN 2024abfo, the He-rich progenitors of SNe IIb and finally the H-free stars generating SNe Ib.

In conclusion, SN 2024abfo is a peculiar SN IIb, with a faint, just noticeable first peak, but with evidence of shock cooling, and H-rich spectra. It reached a faint peak luminosity of $M_r=-16.5$~mag 24 days after the explosion, then faded slowly and linearly for at least 2 months. Its progenitor was consistent with a hot YSG, with a more massive H envelope than the typical progenitors of SNe IIb, which explains the persistence of H lines even at late phases.

\section*{Data availability}
The observed optical Sloan, Johnson, ATLAS and \textit{Swift} magnitudes are tabulated in the photometry tables, which are only available in electronic form at the CDS via anonymous ftp to cdsarc.u-strasbg.fr (130.79.128.5) or via http://cdsweb.u-strasbg.fr/cgi-bin/qcat?J/A+A/.
All the spectra are released on the \textsc{WISeREP} interface (\url{https://www.wiserep.org/object/26537}).

\begin{acknowledgements}
\begin{small}
We thank the anonymous referee for the useful comments and suggested corrections to the manuscript.
We thank N. Elias-Rosa for her comments.

AR acknowledges financial support from the GRAWITA Large Program Grant (PI P. D’Avanzo). AR, AP, GV, IS, SB acknowledge financial support from the PRIN-INAF 2022 "Shedding light on the nature of gap transients: from the observations to the models".

SJS acknowledges funding from STFC Grants ST/Y001605/1, ST/X006506/1, ST/T000198/1, a Royal Society Research Professorship and the Hintze Charitable Foundation. 

T.-W.C. acknowledges the Yushan Fellow Program by the Ministry of Education, Taiwan for the financial support (MOE-111-YSFMS-0008-001-P1).

JD acknowledge support by FCT for CENTRA through Project No. UIDB/00099/2020 and under the PhD grant 2023.01333.BD.

T.E.M.B. is funded by Horizon Europe ERC grant no. 101125877.

NSF's NOIRLab (National Optical-Infrared Astronomy Research Laboratory), the US center for ground-based optical-infrared astronomy, operates the Cerro Tololo Inter-American Observatory (CTIO). It is managed by the Association of Universities for Research in Astronomy (AURA) under a cooperative agreement with NSF.

Based on observations collected at the European Organisation for Astronomical Research in the Southern Hemisphere, Chile, as part of ePESSTO+ (the advanced Public ESO Spectroscopic Survey for Transient Objects Survey – PI: Inserra), under ESO program ID 112.25JQ.

Support to ATLAS was provided by NASA grant NN12AR55G.

The Pan-STARRS1 Surveys (PS1) and the PS1 public science archive have been made possible through contributions by the Institute for Astronomy, the University of Hawaii, the Pan-STARRS Project Office, the Max-Planck Society and its participating institutes, the Max Planck Institute for Astronomy, Heidelberg and the Max Planck Institute for Extraterrestrial Physics, Garching, The Johns Hopkins University, Durham University, the University of Edinburgh, the Queen's University Belfast, the Harvard-Smithsonian Center for Astrophysics, the Las Cumbres Observatory Global Telescope Network Incorporated, the National Central University of Taiwan, the Space Telescope Science Institute, the National Aeronautics and Space Administration under Grant No. NNX08AR22G issued through the Planetary Science Division of the NASA Science Mission Directorate, the National Science Foundation Grant No. AST-1238877, the University of Maryland, Eotvos Lorand University (ELTE), the Los Alamos National Laboratory, and the Gordon and Betty Moore Foundation.

We acknowledge the use of data from the Swift data archive.

\end{small}
\end{acknowledgements}

\bibliographystyle{aa} 
\bibliography{bib}

\begin{appendix} 

\onecolumn 
\FloatBarrier 
\section{Complementary tables}\label{Appendix}

\begin{table*}[h] 
\caption{Observational facilities and instrumentation used in the photometric follow-up of SN 2024abfo.}
\label{tab1}
\begin{tabular}{llll}
\hline
Telescope & Location & Instrument & Filters \\
\hline
\textit{Swift} (0.3m) & Space & UVOT & $UV$ filters+$UBV$ \\
PROMPT (0.4m+0.6m) & CTIO   & Apogee    & $BVgriz$ \\
SLT (0.4m)      & Taiwan    & ANDOR     & $griz$ \\
ATLAS (0.5m)    & Hawaii    & ACAM1     & $c,o$ \\
LCO (1.0m)      & Australia, Chile, South Africa & Sinistro  & $uz$ \\
PS1 (1.8m)      & Hawaii    & GPC1      & $iz$ \\
NTT (3.58m)     & La Silla  & EFOCS2    & $V$ \\
\hline
\end{tabular}
\end{table*}

\begin{table*}[h]
\caption{Log of the spectroscopic observations of SN~2024abfo.}
\label{tab:spectra}
\begin{tabular}{llllll}
\hline
Date & MJD & Phase & Spectral & Resolution & Telescope + Instrument + Grism \\
 & & (d) & range (\AA) & (\AA) & \\
\hline
2024-11-15 & 60630.8 & +2.5 & 4100-7970  & $R=172$ & SAAO 1.0m Lesedi + Mookodi \\
2024-11-20 & 60635.0 & +6.7 & 3650-10000 & 13 & NTT 3.58m + EFOSC2 + gr11/16 \\
2024-11-28 & 60643.2 & +15 & 3370-10000 & 14 & NTT 3.58m + EFOSC2 + gr11/16 \\
2024-12-06 & 60651.2 & +23 & 3370-10000 & 14 & NTT 3.58m + EFOSC2 + gr11/16 \\
2024-12-20 & 60665.2 & +37 & 3370-10000 & 14 & NTT 3.58m + EFOSC2 + gr11/16\\
2024-12-21 & 60666.2 & +38 & 4740-6750  & 6  & NTT 3.58m + EFOSC2 + gr18 \\
2024-12-29 & 60674.2 & +46 & 3370-10000 & 14 & NTT 3.58m + EFOSC2 + gr11/16\\
2025-01-10 & 60686.2 & +58 & 3360-10000 & 13 & NTT 3.58m + EFOSC2 + gr11/16\\
2025-02-09 & 60716.1 & +88 & 3360-10000 & 13 & NTT 3.58m + EFOSC2 + gr11/16\\
\hline
\end{tabular}
\tablefoot{The phases reported are relative to the assumed explosion time (MJD 60628.28).}
\end{table*}

\begin{table}[h]
\caption{Evolution of the velocity of the $H\alpha$, \hei $\lambda$5876 and \Feii~$\lambda$5169 lines. Velocities are expressed in \kms.}
\label{tab:vel_spectra}
\begin{tabular}{cccc}
\hline
Phase (d) & \Ha & \hei $\lambda$5876 & \Feii~$\lambda$5169  \\
\hline
+3  &	22,500$\pm$1,000 & -			  &	-             \\
+7  &	17,500$\pm$500 & 13,500$\pm$1,500 & 10,100$\pm$400 \\
+15 &	14,600$\pm$400 & 8,500$\pm$500 & 9,500$\pm$500 \\
+23 &	13,100$\pm$400 & 8,000$\pm$500 & 8,200$\pm$600 \\
+37 &	12,900$\pm$400 & 7,500$\pm$300 & 5,500$\pm$300 \\
+46 &	12,300$\pm$200 & 7,200$\pm$300 & 5,300$\pm$300 \\
+58 &	11,800$\pm$200 & 7,200$\pm$300 & 4,800$\pm$300 \\
+88 &   10,700$\pm$300 & 7,200$\pm$400 & 4,850$\pm$200 \\
\hline
\end{tabular}
\end{table}

\begin{table*}[h]\centering
\caption{Properties of the identified or modelled progenitors of SNe IIb. SN 1987A is also added for comparison.}
\label{tab:progenitors}
\adjustbox{max width=1.02\textwidth}{
\begin{tabular}{llllllllll}
\hline
Object & Progenitor & $log(T_{\rm eff}/K)$ & $log(L/$\Lsun$)$ & Radius & Mass    & SN has    & Peak abs. & Reference \\
    & detected   &                  &                & (\Rsun)  & (\Msun) & 1st peak? & Mag ($r/R$) & \\
\hline
SN 1987A    & Yes (BSG) & 4.20$\pm$0.04 & 5.04$\pm$0.10 & 45$\pm$15 & $\sim$20 & - & $-16.4$ & \cite{Podsiadlowski1992PASP..104..717P} \\
\hline
SN 1993J    & Yes (RSG) & 3.63$\pm$0.05 & 5.1$\pm$0.3   & 600 & 15         & Yes   & $-17.6$ & \cite{Maund2004Natur.427..129M} \\
SN 2008ax   & Yes (BSG) & 4.05$^{+0.25}_{-0.17}$ & 4.69$^{+0.61}_{-0.27}$  & 50  & <25       & No    & $-18.0$   & \cite{Folatelli2015ApJ...811..147F} \\
SN 2011dh   & Yes (YSG) & 3.78$\pm$0.02 & 4.92$\pm$0.20 & 200 & 13$\pm$3   & No?    & $-17.2$ & \cite{Maund2011ApJ...739L..37M} \\
SN 2011fu   & No  & -             & -             & 450 & 13-18      & Yes   & $-18.1$ & \cite{Morales-Garoffolo2015} \\
SN 2011hs   & No  & -             & -             & 500-600 & 12-15  & No    & $-16.9$   & \cite{Bufano2014MNRAS.439.1807B} \\
SN 2013df   & Yes (RSG) & 3.63$\pm$0.01 & 4.94$\pm$0.06 & 600 & 13-17      & Yes   & $-17.8$ & \cite{VanDyk2014AJ....147...37V} \\
SN 2016gkg  & Yes (BSG) & 3.98$^{+0.21}_{-0.16}$ & 5.14$^{+0.22}_{-0.39}$  & 140$^{+130}_{-100}$ & 15-20      & Yes   & $-17.2$   & \cite{Kilpatrick2017MNRAS.465.4650K} \\
SN 2017gkk  & Yes (YSG) & 3.72$\pm$0.08 & 5.17$\pm$0.04 & -   & 16         & -     & -       & \cite{Niu2024ApJ...970L...9N} \\
SN 2024abfo & Yes (YSG) & 3.80-3.85 & 5.00$\pm$0.05 & 215-270 & 15 & faint & $-16.5$ & This work \\
\hline
\end{tabular}
}
\tablefoot{If the progenitor was not detected in archival images its parameters are inferred from hydrodynamical modelling. The masses reported are the initial masses of the progenitors.
We note that an early peak might be visible in the $g$-band light curve of SN 2011dh published by \cite{Arcavi2011ApJ...742L..18A}.
}
\end{table*}

\end{appendix}

\label{LastPage}
\end{document}